\documentclass[12pt]{article}

\def \pl {\partial}
\def \lf {\left (}
\def \rt {\right )}

\def \beq {\begin{equation}}
\def \eeq {\end{equation}}
\def \To {\Rightarrow}

\begin{document}

\begin{titlepage}

\begin{center}

{\hbox to \hsize{\hfill hep-th/0204107}}
{\hbox to \hsize{\hfill CU-TP-1054}}

\bigskip

\vspace{6\baselineskip}

{\Large \bf New Coordinates for de Sitter Space and de Sitter Radiation}

\bigskip
\bigskip
\bigskip

{\large \sc Maulik K. Parikh{\small} 
\footnote{e-mail: {\tt mkp@phys.columbia.edu}}\\[1mm]}

{\em Department of Physics, Columbia University, New York, NY 10027}\\[3mm]

\vspace*{1.5cm}
\large{

{\bf Abstract}\\

}
\end{center}
\noindent
We introduce a simple coordinate system covering half of de Sitter space. 
The new coordinates have several attractive properties: the time
direction is a Killing vector, the metric is smooth at the horizon,
and constant-time slices are just flat Euclidean space. We demonstrate
the usefulness of the coordinates by calculating the rate at which
particles tunnel across the horizon. When self-gravitation is taken
into account, the resulting tunneling rate is only approximately
thermal. The effective temperature decreases through the emission of
radiation.

\end{titlepage}

\newpage

\setcounter{page}{2}

\section{Motivation}
At the heart of Einstein's theory of gravity is local diffeomorphism
invariance: coordinate systems are unimportant, only diffeomorphism
invariants matter. However, a poor choice of local coordinates can
sometimes obscure the nature of the global aspects of spacetime, such
as horizons or causal boundaries. Indeed, the true nature of the
``coordinate singularity'' at the Schwarzschild radius eluded Einstein
himself, and was only fully illuminated with the discovery of
coordinate systems that were regular at the horizon. Coordinate
systems that cover larger patches of spacetime are especially
useful to have in dealing with physical phenomena that are in some
sense nonlocalized.

In this note, we present a simple new coordinate system for de Sitter
space, covering the causal future/past of an observer. The new
coordinates, which we shall call Painlev\'e-de Sitter coordinates, are
a cross between static coordinates and planar coordinates, and inherit
the strengths of each of these. For example, like static coordinates
but unlike planar coordinates, the new coordinates have a direction of
time that is a Killing vector, making them well-adapted to
thermodynamics. On the other hand, like planar coordinates, but unlike
static coordinates, Painlev\'e-de Sitter coordinates continue smoothly
through the horizon, and constant time slices are just flat Euclidean
space. Painlev\'e-de Sitter coordinates differ also from
Eddington-Finkelstein type coordinates in that the coordinates are all
either timelike or spacelike, rather than null.

The combination of a Killing time direction and regularity at the
horizon is particularly powerful as it allows one to study
across-horizon physics as seen by an observer. A natural application
is de Sitter radiation. Heuristically, one envisions de Sitter
radiance as arising in much the same way that Hawking radiation
does. That is to say, a particle-pair forms just inside the horizon,
one member of the pair tunnels across the horizon, and the virtual
pair becomes real. To show that this is actually what happens, one
would like to compute the amplitude for traversing the horizon;
because of their regularity, Painlev\'e-de Sitter coordinates make the
calculation feasible. By directly evaluating the imaginary part of the
action, we obtain the emission amplitude associated with a tunneling
particle. In the s-wave limit, it is in fact possible to extend the
computation to include the effects of self-gravitation. As a result,
the de Sitter spectrum turns out to be only approximately thermal. In
particular it has a cutoff at high energies. The back-reaction is such
that, unlike Schwarzschild black holes, de Sitter space lowers its
temperature as it radiates.

\section{Painlev\'e-de Sitter Coordinates}
Many different coordinate systems are known for de Sitter space (see,
e.g., \cite{hawkingellis,dsreview}). One commonly used metric is the
static metric, analogous to the familiar Schwarzschild metric for an
uncharged black hole. This metric covers a static patch, that part of
de Sitter space that an observer at the origin can interact with. The
time coordinate, $t_s$, corresponds to a timelike Killing vector,
which makes it suitable for thermodynamics; thermal equilibrium
requires among other things that the spatial metric be at
equilibrium. However, the static metric has the limitation that it is
only valid upto the horizon; it therefore covers only a very small
region of the full space. Another drawback is that in a static
background one cannot expect to get radiation, a phenomenon that is
manifestly time-reversal asymmetric. Indeed, in early calculations of
Hawking radiation from black holes, time-reversal symmetry had to be
broken by hand through the introduction of a collapsing surface.

We shall now illustrate the method for obtaining Painlev\'e-type coordinates,
valid across the horizon. Consider then a general static metric of the form
\beq
ds^2 = - (1 -g(r)) dt_s^2 + {dr^2 \over 1 - g(r)} + r^2 d
\Omega_{D-2}^2 \; .
\eeq
Here $g(r) = 1$ corresponds to a horizon; we shall assume for simplicity
that there is only one horizon. For de Sitter space, $g(r) = r^2/l^2$
and $r = l$ marks the horizon. $r = 0$ could be the worldline of an
observer at the origin.

To obtain the new line element, define a new time coordinate, $t$, by
$t_s = t + f(r)$. The function $f$ is required to depend only on $r$
and not $t$, so that the metric remains stationary, i.e.
time-translation invariant. Stationarity of the metric automatically
implements the desirable property that the time direction be a Killing
vector. What other conditions should we impose on $f$? Our key
requirement is that the metric be regular at the horizon. We can
implement this as follows. We know that a radially free-falling
observer who falls through the horizon does not detect anything
abnormal there; we can therefore choose as our time coordinate the
proper time of such an observer. As a corollary, we demand that
constant-time slices be flat. We then obtain the condition
\beq
{1 \over 1 - g(r)} - (1-g(r))(f'(r))^2 = 1 \; .	\label{fdiffeqn}
\eeq
For de Sitter space, $g(r) = r^2/l^2$, and the solution to Eq. (\ref{fdiffeqn})
yields 
\beq
t_s = t \pm {l \over 2} \ln (1 - r^2/l^2 ) \; .	\label{tsandt}
\eeq
As usual, the fact that the transformation is singular at $r = l$
merely indicates that the original coordinate $t_s$ was ill-defined
there. Choosing the minus sign for now, gives us the new coordinates,
which we shall call Painlev\'e-de Sitter coordinates. The desired line
element is 
\beq
\fbox{$
ds^2 = - \lf 1 -{r^2 \over l^2} 
\rt dt^2 - 2 {r \over l} dt~ dr + dr^2 + r^2 d \Omega^2 \; .
$}
\eeq
The Painlev\'e-de Sitter metric has a number of attractive features.
First, none of the components of either the metric or the inverse
metric diverge at the horizon. Second, by construction constant-time
slices are just flat Euclidean space. Third, the generator of $t$ is a
Killing vector. ``Time'' becomes spacelike across the horizon, but is
nevertheless Killing; this fact can be exploited to compute global
charges such as mass \cite{klemm,mass} in a natural way. Finally, an
observer precisely at the origin does not make any distinction between
these coordinates and static coordinates; indeed, the function $f$
that distinguishes the two time coordinates vanishes there. 

We call this the Painlev\'e-de Sitter metric because the analogous
line element for four-dimensional Schwarzschild black holes is
\beq
ds^2 = - \lf 1 - {2 M \over r} \rt dt^2 - 2 \sqrt{2 M \over r}
dt~ dr + dr^2 + r^2 d \Omega_2^2 \; .
\eeq
This superb though relatively unknown line element was discovered by
Painlev\'e \cite{painleve} many years ago; it was rediscovered in a
modern context by Kraus and Wilczek \cite{per}. Similar coordinate
systems have been found for black holes in anti-de Sitter space
\cite{ads,vagenas}.

These coordinates cover the observer's causal future, or half of the
full space. The metric is necessarily nonstatic (that is, not
time-reversal invariant) as the causal patch is itself not
time-reversal invariant. To obtain the other half of the space,
corresponding to the causal past of an observer at the antipode, one
chooses the opposite (plus) sign for the off-diagonal component {\em
and} reverses the sense of time, so that $t$ increases to the
past. Choosing the other sign in Eq. (\ref{tsandt}) without reversing
the direction of time gives a metric that covers the causal past of
the original observer. The relation between Painlev\'e-de Sitter
coordinates and planar coordinates is made clear by the transformation
\beq 
r = \rho e^{t/l} \To ds^2 = -dt^2 + e^{2t/l} \lf d \rho^2 +
\rho^2 d \Omega^2_{D-2} \rt \; .  
\eeq 
We see that the new coordinates have the same radial coordinate as
static coordinates and the same time coordinate as planar coordinates.

We conclude this section by writing down the radial geodesics in these
coordinates. The null radial geodesics obey 
\beq
{dr \over dt} = r/l \pm 1 \; ,	\label{null}
\eeq
where the plus (minus) sign corresponds to rays that go away from (towards)
the observer. When the particle is beyond the horizon ($r>l$), both
ingoing and outgoing trajectories correspond to increasing $r$, and the
particle cannot (classically) cross the horizon. The general solution is
\beq
r(t) = l \lf e^{ t / l } \mp 1 \rt \; .
\eeq
Radially-directed null rays leaving the observer at $t = 0$ reach the
horizon at $t =  l \ln 2$, and reach future null infinity at $t = \infty$.
Turning now to massive particles, the radial geodesic equation implies
\beq
U^r = {dr \over d \tau} = \lf 1 - r^2/l^2 \rt m^2 p_t \; ,
\eeq
where $\tau$ is the proper time. Here (by stationarity) $-p_t$ is a
constant of the motion, being equal to the energy measured by an
observer at $r=0$. Note that if we set $p_t (= m U_t) \equiv -m$, then
using $U^2 = -1$ we find that
\beq
U^t = 1 \To \tau = t + c \; ,
\eeq
so the Painlev\'e-de Sitter time coordinate is nothing more than the
proper time along a radial geodesic worldline, such as that of a
free-falling observer. Indeed, that is precisely how we arrived at
these coordinates in the first place.

\section{Tunneling Across the de Sitter Horizon}
The great utility of having a coordinate system that is well-behaved
at the horizon is that one can study across-horizon physics. In
this section, we will determine the temperature of de Sitter space by
directly computing the rate at which particles tunnel across the
horizon. Our computation will parallel the analogous calculation for
black holes, in which Hawking radiation is expressed as a tunneling
phenomenon \cite{tunnel}. 

Now, because of the infinite blueshift near the horizon, the
characteristic wavelength of any wavepacket is always arbitrarily
small there, so that the geometrical optics limit becomes an
especially reliable approximation. This is of course a big plus: the
geometrical optics limit allows us to obtain rigorous results directly
in the language of particles, rather than having to use the unwieldy
and physically less transparent Bogolubov method that is more
traditionally used. Moreover, as we shall see, the inclusion of
back-reaction effects is also perhaps easier now.

In any event, since we are in the semi-classical limit, we can
apply the WKB formula. This relates the tunneling amplitude to the
imaginary part of the particle action at stationary phase. (The phase
is $i \int L dt / \hbar = i \int (p \dot{x} - H) dt / \hbar$; since
energy is real, exponential damping comes from the imaginary part of
emission rate $\int p dx$, which in the nonrelativistic limit becomes
the usual $\int dx \sqrt{2m(V(x)-E)}$. The imaginary part of the action,
$I$, is thus given by the imaginary part of the momentum integral.)
The emission rate, $\Gamma$, is the square of the tunneling amplitude: 
\beq
\Gamma \sim \exp (- 2 ~{\rm Im}~ I/ \hbar) \approx \exp (-\beta E) \; .
\label{gamma}
\eeq
On the right-hand side, we have equated the emission probability to
the Boltzmann factor for a particle of energy $E$. To the extent
that the exponent depends linearly on the energy, the thermal
approximation is justified; we can then identify the inverse
temperature as the coefficient $\beta$.

We will consider here the s-wave emission of massless
particles. Higher partial wave emission is in any case suppressed by
$\hbar$. In the s-wave, particles are really massless shells. If we
imagine a shell to consist of constituent massless particles each of
which travels on a radial geodesic, then we see that the motion of the
shell itself must follow the radial null geodesic for a particle. That
is, it obeys Eq. (\ref{null}), with the minus sign. We will use these
radial geodesics to compute the imaginary part of the action, as
follows.

Since the calculation involves a few tricks \cite{tunnel}, we outline
it here before putting in the details. First observe that we can
formally write the action as 
\beq
{\rm Im}~ I = {\rm Im}~ \int_{r_i}^{r_f} p_r ~dr = {\rm Im}~ \int_{r_i}^{r_f}
\int_0^{p_r} dp'_r ~dr \; ,
\eeq
where $p_r$ is the radial momentum. We expect $r_i$ to correspond
roughly to the site of pair-creation, which should be slightly outside
the horizon. (Note that the second member of the pair contributes
nothing to the tunneling rate, since it is always classically allowed
and therefore has real action.) We expect $r_f$ to be a classical
turning point, at which the semi-classical trajectory (i.e. instanton)
can join onto a classical-allowed motion. This must be slightly within
the horizon, else the particle would not be able to propagate
classically from there to the observer. However, the precise limits on
the radial integral are unimportant, so long as the range of
integration includes the horizon. 

We now eliminate the momentum in favor of energy by using Hamilton's equation
\beq
\left . {d H \over dp} \right |_r = {\pl H \over \pl p} = {dr \over
dt} \; ,
\eeq
where the Hamiltonian, $H$, is the generator of Painlev\'e time. Hence
within the integral over $r$, one can trade $dp$ for $dH$. Without
being very careful about signs, the integral over $H$ now just gives the
particle energy $E$. However, substituting Eq. (\ref{null})
for a particle going radially towards $r = 0$, the radial integral has
a simple pole at the horizon: 
\beq
{\rm Im}~ I = {\rm Im}~ \int_{r_i}^{r_f} \int_0^E {dH \over {dr
\over dt}} dr = {\rm Im}~ E \int_{r_i}^{r_f} {l dr \over r - l} \; .
\eeq
The pole lies along the line of integration, and therefore yields $\pi
i$ (rather than $2 \pi i$) times the residue. Again, we postpone
consideration of the sign associated with the direction of the
contour. We get
\beq
{\rm Im}~ I = \pi l E \; .
\eeq
Consulting Eq. (\ref{gamma}), we find that this corresponds to a temperature
\beq
T_{dS} = {\hbar \over 2 \pi l}	\label{temp} \; ,
\eeq
which is precisely the temperature of de Sitter space. To summarize:
Painlev\'e-de Sitter coordinates have allowed us to compute the
radiation rate directly from the particle action, with the action
incurring an imaginary part from a pole at the horizon.

Let us now do the calculation more carefully, keeping track of the
signs, and including the effects of back-reaction. Perhaps it should
be stressed that the reason we are interested in back-reaction is not
merely to compute higher order in $E$ effects, but because
self-gravitation is central to the entire process of across-horizon
tunneling. Without self-gravitation the back-of-the-envelope
calculation above is puzzling: if this is tunneling, where is the
barrier? Put another way, if particles created just inside the horizon
have only to tunnel just across -- an infinitesimal separation -- what
characterizes the scale of the tunneling? Recall that in the Schwinger
process of electron-positron pair production in an electric field,
there is a nonzero separation scale, $r \sim mc^2/qE$, between the
classically allowed configurations. In the following, we will see
that, as with Hawking radiation \cite{tunnel,alexey}, self-gravitation
resolves these issues. Back-reaction results in a shift of the horizon
radius; the finite separation between the initial and final radius is
the classically-forbidden region, the barrier. 

How does one incorporate back-reaction? In a general situation, this
is a notoriously difficult problem, calling for a theory of quantum
gravity. Indeed, generically one has to worry about how to
consistently quantize the gravitational waves produced by a matter
source. However, for the special case of spherical gravity it is
possible to integrate out gravity, at least semi-classically. This is
because for spherical gravity, Birkhoff's theorem (more precisely, its
generalization to a nonzero cosmological constant) states that the
only effect on the geometry that the presence of a spherical shell
has, is to provide a junction condition for matching the total mass
inside and outside the shell. (In three dimensions, where there are no
gravitational waves, it may be possible to compute the emission rates
for the higher partial waves as well.)

Since the geometry is different on the two sides of the shell, one can
now ask which geometry determines the motion of the shell. (Thus,
self-gravitation automatically breaks the principle of equivalence.)
The geometry inside the shell is empty de Sitter space, while the
geometry outside is that of Schwarzschild-de Sitter space with energy
$E$. (Technically, empty de Sitter space has a mass too \cite{mass};
what follows is unaffected by this shift.) It is the outside
$E$-dependent metric that determines the motion of the
self-gravitating shell. Consider, for simplicity, $dS_3$;
generalization to higher dimensions is straightforward. The effective
geometry whose radial geodesic determines the motion of the shell has
the line element
\beq 
ds_{\rm effective}^2 = - (1 - 8 G E - r^2/l^2) dt_s^2 + {dr^2 \over 1 - 8 G E
- r^2/l^2} + r^2 d \phi^2 \; .  
\eeq 
Here we have inserted Newton's constant, and $E > 0$ is physically the
energy of the shell as measured by an observer at $r = 0$. The
corresponding Painlev\'e metric is now 
\beq 
ds_{\rm effective}^2 = - (1 - 8 G E - r^2/l^2) dt^2 - 2 \sqrt{r^2/l^2
+ 8 G E} ~ dt~ dr + dr^2 + r^2 d \phi^2 \; . 
\eeq
The imaginary part of the action is then 
\beq 
{\rm Im}~I = {\rm Im}~ \int_0^H \int_{r_i}^{r_f} {dr ~ dH' \over
\sqrt{r^2/l^2 + 8G E'} - 1} \; , 
\eeq 
where we have inserted the radial geodesic derived from the
effective metric. Here $r_i = l$ is the original radius of the
horizon just before pair-creation, while $r_f$ is the {\em new} radius
of the horizon, and is equal to $l \sqrt{1 - 8 G E}$. What matters is
that $r_f < r_i$. The Feynman prescription for evaluating the sign of
the contour is to displace the energy from $E'$ to $E' - i
\varepsilon$. Substituting $u = r^2$, 
\beq 
{\rm Im}~I = {\rm Im}~ \int_0^H \int_{u_i}^{u_f} {du \over 2 \sqrt{u}}
{l \sqrt{u + 8 G l^2 ( E' - i \varepsilon )} + l^2 \over
u - (l^2  - 8 G l^2 (E' - i \varepsilon))} ~ dH' \; , 
\eeq
we see that the pole lies in the upper-half $u$-plane. Doing the $u$
integral first we find
\beq 
{\rm Im}~ I = - \pi l \int_0^H {d H' \over \sqrt{1 - 8 G E'}} \; .  
\eeq
Now the total energy of de Sitter space {\em decreases} when
positive-energy matter is added to it \cite{mass}, because of the
negative gravitational binding energy. Therefore the Hamiltonian $H$
satisfies $d H = -dE$, giving
\beq 
{\rm Im}~ I = - {\pi l \over 4G} \lf \sqrt{1 - 8 G E} - 1 \rt \; , 
\eeq 
and the tunneling rate is therefore
\beq 
\fbox{$ \Gamma \sim \exp \lf +{\pi l \over 2G \hbar} \lf \sqrt{1 - 8 G
E}-1 \rt \rt \; . $} \label{rate}
\eeq
When the particle's energy is small, $8 G E \ll 1$, the square root
can be approximated. To linear order in $G E$, we recover our
previous back-of-the-envelope result, Eq. (\ref{temp}). As a check, we
note that the sign has also come out correctly. To this order then, the
thermal approximation is a good one. But at higher energies the
spectrum cannot be approximated as thermal. Indeed, the spectrum has
an ultraviolet cutoff at $8 G E = 1$, beyond which there is no radiation
whatsoever. The precise expression, Eq. (\ref{rate}), is related to
the change in de Sitter entropy. The entropy of three-dimensional de
Sitter space is
\beq
S = {2 \pi r_H \over 4 G \hbar} \; ,
\eeq
where $r_H$ is the horizon radius. Once the shell has been emitted,
its energy causes the horizon to shrink to the new radius $r_H = l
\sqrt{1 - 8G E}$. Thus, Eq. (\ref{rate}), can be written as the
exponent of the difference in the entropies, $\Delta S$, before and
after the particle has been emitted. This also explains the UV cutoff:
the horizon cannot shrink past zero. Incidentally, Eq. (\ref{rate})
takes the same form as the corresponding expression for Hawking
radiation from a Schwarzschild black hole \cite{tunnel,thesis}, once
back-reaction effects have been taken into account.

Indeed, one expects Eq. (\ref{rate}) on general grounds. For whatever the
ultimate form of the holographic description of de Sitter space,
quantum field theory tells us (via Fermi's Golden Rule) that the
rate must be expressible as 
\beq
\Gamma (i \to f) = |{\cal M}_{fi}|^2 
\cdot \lf \mbox{\rm phase space factor} \rt \; ,
\eeq
where the first term on the right is the square of the amplitude 
for the process. The phase space factor is obtained by summing over
final states and averaging over initial states. But the number of final
states is just the final exponent of the final entropy, while the number of
initial states is the exponent of the initial entropy. Hence
\beq
\Gamma \sim {e^{S_{\rm final}} \over e^{S_{\rm initial}}} = \exp
(\Delta S) \; 
\eeq
in agreement with our result.

We end this section by noting that, although we have derived de Sitter
radiance directly as particles tunneling across the de Sitter horizon,
an alternate viewpoint is also possible. In this view, there is no de
Sitter horizon and the spacetime is cutoff by a membrane living at the
horizon. An observer then interprets the radiation not as tunneling
particles, but rather as the spontaneous emissions of the membrane. In
\cite{membrane,thesis} it was shown that the classical equations of motion
of the membrane can be derived from an action, and that Euclideanizing
this action yields the correct entropy. It would be interesting to see
whether the radiation can also be understood in this language.

\section{On the Thermal Stability of de Sitter Space}
The signs in Eq. (\ref{rate}) have the consequence that
three-dimensional de Sitter space is thermally stable. For after a
particle has been emitted, the new horizon radius is smaller than what
it had previously been. The probability for emission of a second
particle is now 
\beq
\Gamma_2 \sim \exp \Delta S = \exp \lf {\pi l \over 2G \hbar} \lf \sqrt
{1-8G (E_1 + E_2)} - \sqrt{1 - 8 G E_1} \rt \rt \; ,
\eeq
where $E_1$ and $E_2$ are the energies of the two particles. For
small $E_2$, this is 
\beq
\Gamma_2 \approx \exp \lf {-2 \pi l \over \sqrt{1 - 8 G E_1}} {E_2
\over \hbar} \rt \; .
\eeq
The effective temperature which governs the emission probability of
the second particle is therefore
\beq
T_2 = {\hbar \sqrt{1 - 8 G E_1} \over 2 \pi l} \; ,
\eeq
which is {\em lower} than before. This is to be contrasted with the
situation for Schwarzschild black holes, for which there is a runaway
explosion as the black hole becomes smaller. In that respect, de
Sitter space more closely resembles charged black holes. Note also
that the change in the temperature of de Sitter space takes place
because of the matter inside it, and not by any change in the cosmological
constant, which of course remains constant throughout.

The decrease in the effective temperature holds out the possibility of
thermally stabilizing de Sitter space \cite{abbottdeser}. We started
by considering empty de Sitter space. As this radiates and lowers its
temperature, the horizon volume fills up with radiation. Eventually,
the radiation passes through the origin and leaves the horizon; this
causes the horizon radius to increase again, raising the
temperature. By detailed balance, a stable thermodynamic equilibrium
can be reached.

In higher dimensions ($D > 3$), the story is somewhat different. The
presence of the de Sitter radiation can now lead to the formation of a
black hole, which in turn will radiate Hawking radiation
outwards. Equilibrium is presumably reached only when the de Sitter
and black hole horizons are coincident.

\begin{flushleft}
{\sc Acknowledgments}
\end{flushleft}

\noindent
It is a pleasure to thank Daniel Kabat, Ivo Savonije, and especially Erik
Verlinde for discussions. The author is supported by DOE grant
DF-FCO2-94ER40818.

\end{document}